# Cloud Computing Benefits for Educational Institutions

Mr. Ramkumar Lakshminarayanan[1], Dr. Binod Kumar[2], Mr. M. Raju[3]
Higher College of Technology, Muscat, Oman
rajaramcomputers@gmail.com[1], binod.kumar@hct.edu.om[2],
marappan.raju@hotmail.com[3]

**ABSTRACT**

Education today is becoming completely associated with the Information Technology on the content delivery, communication and collaboration. The need for servers, storage and software are highly demanding in the universities, colleges and schools. Cloud Computing is an Internet based computing, whereby shared resources, software and information, are provided to computers and devices on-demand, like the electricity grid. Currently, IaaS (Infrastructure as a Service), PaaS (Platform as a Service) and SaaS (Software as a Service) are used as business model for Cloud Computing. The paper also introduces the cloud computing infrastructure provided by Microsoft, Google and Amazon Web Service. In this paper we will review the features the educational institutions can use from the cloud computing providers to increase the benefits of students and teachers.

**Keywords**: IaaS, PaaS, SaaS, Amazon, Microsoft Live@edu, Google Apps.

## I. INTRODUCTION

To achieve human goals one of the prerequisite is education. From various researches it is clear that the human welfare developments are associated with Information and Communication Technologies commonly known as ICT [2]. A relevant education is more important today than ever, because today's Networked World demands a workforce that understands how to use technology as a tool to increase productivity and creativity. With the demand for the needs of infrastructure, software and platform the need for a technology is required by the institution.

Cloud computing is a model for enabling convenient, on-demand network access to a shared pool of configurable computing resources (e.g., networks, servers, storage, applications, and services) that can be rapidly provisioned and released with minimal management effort or service provider interaction. This cloud model promotes availability and is composed of five essential characteristics, three service models, and four deployment models [1]. The usage of information technology by the universities, colleges and schools for imparting the training programs are gradually increasing. The need for the networks, servers, storage, applications and services are drastically growing. Educational Institutions have started investing on the infrastructure, platform and software. Educational Institutions demand for the computing needs keep on changing from time to time. The student expectation is to view the information in his PDA, Tablets and Mobile Phones. A solution which maps the needs of educational institution is Cloud Computing. The Characteristics of Cloud Computing is on-demand Self-Service, Broad Network Access, Resource Pooling, Rapid Elasticity and Measured Service.

The Service Models are Cloud Software as a Service, Cloud Platform as a Service and Cloud Infrastructure as a Service. The deployment models are Private Cloud, Community Cloud, Public Cloud and Hybrid Cloud [1].

## II. CLOUD: OVERVIEW

Cloud Computing can be defined as new style of computing in which dynamically scalable and often virtualized resources are provided as a service over the Internet. Cloud Computing has become a significant technology trend, and many experts expect the cloud computing will reshape Information Technology (IT) processes and the IT marketplace. With the cloud computing technology, users uses a variety of devices, including PCs, laptops, smartphones, and PDA's to access programs, storage and application-development platforms over the Internet, via services offered by cloud computing providers. Advantages of the cloud computing technology include cost savings, high availability and easy scalability [3]. Cloud options range from everyday services, such as email, calendaring and collaboration tools that members can collaborate online. System Administrators can bring new services and computing capacity online quickly while managing costs as operational expenses. By allowing IT to respond quickly to changes, cloud computing helps administration manage risks, peak demand, and long term planning needs.

*A. Service Model*

The offering of cloud is in three different models viz., Infrastructure as a Service (IaaS), Platform as a Service (PaaS), and Software as a Service (SaaS).

Infrastructure as a Service (IaaS). The IT infrastructures like processing, storage, networks and other fundamental computing resources can be used by the consumers as a service. In order to integrate/decompose physical resources IaaS uses Virtualization extensively. Amazaon EC2 is an example for IaaS.

Platform as a Service (PaaS). To develop cloud services and applications PaaS provides a development platform supporting the full "Software Lifecycle". PaaS requires programming environment, tools, configuration management etc., to support the application hosting environment. Google App Engine is the example for Platform as a Service.

Software as a Service (SaaS). The Software usage is provided to a consumer as a Service. Based on the demand the consumer can choose his software to use. Cloud providers release their applications on a hosting environment, which can be accessed through networks from various clients like web browser, PDA, etc., by the application users. SalesForce.com, Google Mail, Google Docs are examples for SaaS.

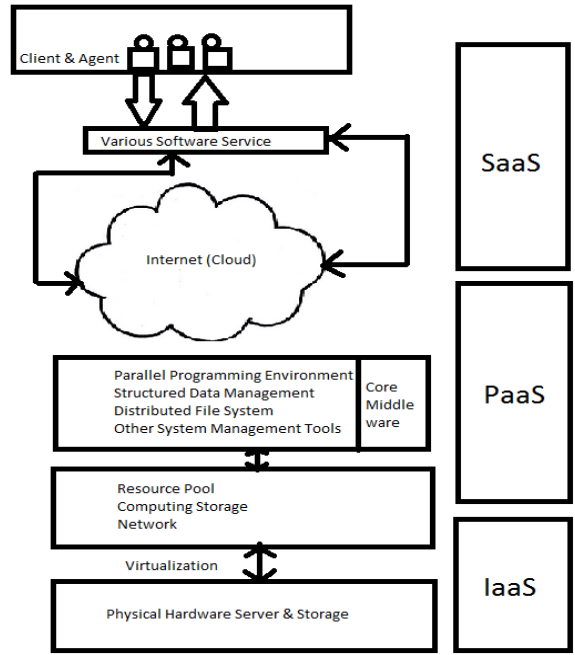

**Service Model**

*B. Deployment Model*

The deployment models defined by the cloud community are Public Cloud, Private Cloud, Hybrid Cloud and Community Cloud.

Public Cloud: One of the leading forms of the current computing deployment model. Mainly used by the general public cloud consumer and the policy, value and costing are defined by the service provider. The popular public cloud services are Amazon EC2, S3, Google App Engine, and Force.com.

Private Cloud: This is a cloud model for a single organization and managed by organization or a third party. The infrastructure can be located on premise or off premise. Primary reason for implementing private cloud is to maximize and utilize existing in-house resources. Secondary reasons include the data privacy and trust for security. Finally, data transfer cost and to have full control over mission-critical activities behind the firewalls. Academic institutions build private cloud for research and teaching purpose.

Hybrid Cloud: It is a combination of two or more clouds viz., private, community or public. In order to optimize the resource and to utilize core competency of the public cloud organizations use the hybrid cloud. Virtual Private Cloud (VPC) is a deployment model of Amazon Web Services (AWS). Using VPC it is possible to have a secure and seamless bridge between IT infrastructure of an organization and Amazon public cloud. Hybrid cloud is the combination of public and private cloud.

Community Cloud: Several organization of same group shares their cloud resources and jointly constructs the policies and requirements. The infrastructure of the cloud can be hosted by a third-party vendor or within one of the organizations in the community.

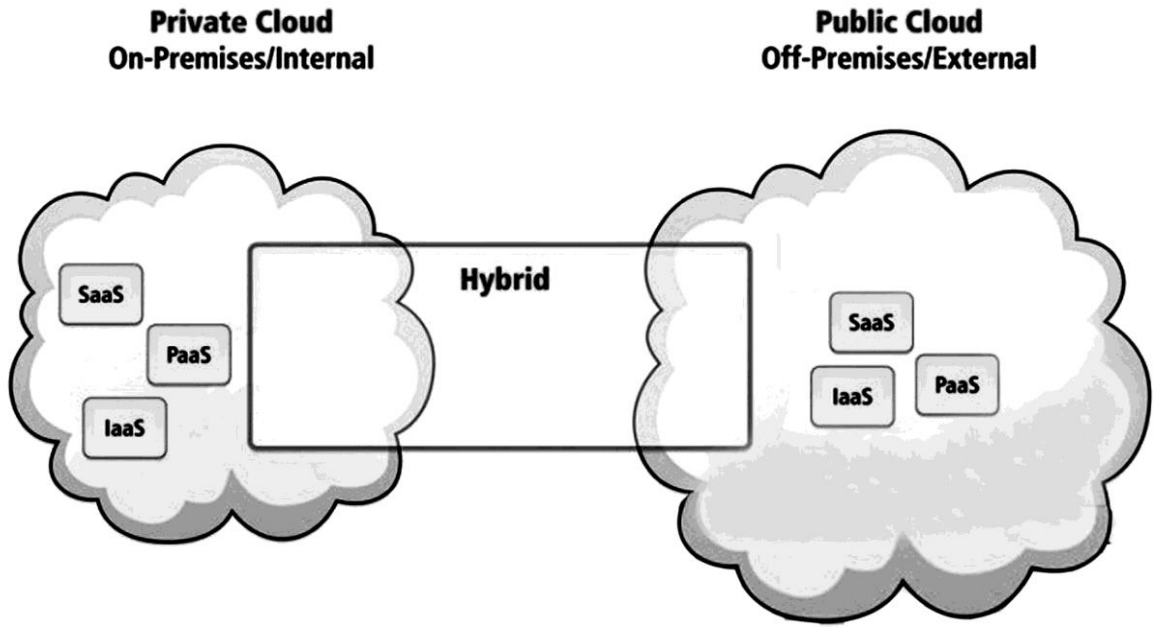

**Cloud Computing Deployment Model**

## III. CLOUD COMPUTING PROVIDERS

*A. Microsoft Live@edu for education*

Microsoft Live@edu is intended for educational needs. It provides a set of hosted collaboration services for the educations institutions. The hosted service includes collaboration services, communication tools, mobile, desktop, and web-based applications. It has the feature of data storage capabilities. Office Live Workspace, Windows Live SkyDrive, Windows Live Spaces, Microsoft Shared View Beta, Microsoft Outlook Live, Windows Live Messenger and Windows Live Alerts are the part of Live@edu suite. By means of free registration process universities, colleges and schools can enroll in the program [4]. Microsoft Live@edu is mainly for the institutions for enabling facilities for their academic activities.

*B. Google Apps for Education*

Google Apps is a collection of web-based programs and file storage that run in a web browser, without requiring users to buy or install software. Users can simply log in to the service to access their files and the tools to manipulate them. The communication tools of Google Apps are Gmail, Google Talk, and Google Calendar and the productivity tools are Google Docs: text files, spreadsheets, and presentations, iGoogle and Google Sites to develop web pages [5]. The tools are free, or users can pay for a Premium Edition that adds more storage space and other features. An Education Edition includes most of the extras in the Premium Edition and is offered at no cost to K–12 (*designation for the sum of primary and secondary education and higher education*). Google Apps allows institutions to use their own domain name with the service and to customize the interface to reflect the branding of that institution. In this way, a college or university can offer the functionality of Google Apps in a package.

*C. Amazon Web Services for Education (AWS)*

Amazon Web Services provides the cloud services in categories of Compute, Software, Content Delivery, Database, Storage, Deployment & Management, Application Services and Workforce [6].

Compute service includes Amazon Elastic Computer Cloud (EC2), Amazon Elastic MapReduce, Auto Scaling and Elastic Load Balancing. Amazon Elastic Compute Cloud delivers scalable, pay-as-you-go compute capacity in the cloud. Amazon Elastic MapReduce is a web service that enables businesses, researchers, data analysts, and developers to easily and cost-effectively process vast amounts of data. Auto Scaling allows user to automatically scale your Amazon EC2 capacity up or down according to conditions. Elastic Load Balancing automatically distributes incoming application traffic across multiple Amazon EC2 instances.

In Software, AWS Marketplace is an online store that helps customers find, buy, and immediately start using software that runs on the AWS cloud. It includes software from trusted vendors like SAP, Zend, Microsoft, IBM, Canonical, and 10gen as well as many widely used open source offerings including Wordpress, Drupal, and MediaWiki.

In Content Delivery, Amazon CloudFront is a web service that makes it easy to distribute content with low latency via a global network of edge locations.

In Database, it has the category of Amazon Relational Database Service (RDS), Amazon DynamoDB, Amazon SimpleDB and Amazon Elastic Cache.

Amazon Relational Database Service is a web service that makes it easy to set up, operate, and scale a relational database in the cloud. Amazon DynamoDB is a fully-managed, high performance, NoSQL database service that is easy to set up, operate, and scale. Amazon SimpleDB is a managed NoSQL database service designed for smaller datasets. Amazon ElastiCache is a web service that makes it easy to deploy, operate, and scale an in-memory cache in the cloud.

In Networking, the classifications are Amazon Route S3, Amazon Virtual Private Cloud (VPC) and AWS Direct Connect.

In Storage, depending on the needs the service provided by AWS are Amazon Simple Storage Service(S3), Amazon Glacier, Amazon Elastic Block Store (EBS), AWS Import/Export and AWS Storage Gateway.

Application Services of AWS are Amazon CloudSearch, Amazon Simple Workflow Service (SWF), Amazon Simple Queue Service (SQS), Amazon Simple Notification Service (SNS) and Amazon Simple Email Service (SES).

In Workforce, Amazon Mechanical Turk enables companies to access thousands of global workers on demand and programmatically integrate their work into various business processes.

As for as education, educators, academic researchers, and students can apply to obtain free usage credits and can utilize on-demand infrastructure. With the grants, educational institutions have made advances in research, enable High-Performance Computing and tackled Big Data. AWS is providing educators up to $100USD as grants as free usage for each student enrolled in courses.

Researchers around the world have access to global computing infrastructure and storage capacity of the AWS cloud. Instead of purchasing a large amount of hardware, researchers can get started by simply opening an AWS account. With services like Amazon Elastic MapReduce much of the heavy lifting of provisioning and configuring Hadoop clusters for data-intensive processing is eliminated. The feature is available for the researchers with grants.

AWS in Education is supporting student organizations around the world and compelling entrepreneurial student initiatives including Project Olympus at Carnegie Mellon, Teams in Engineering Service at the University of California, San Diego, and the "3 Day Start Up" event at the University of Texas, Austin. AWS provides Project Grants supporting free usage of AWS to student organizations and student entrepreneurial projects. AWS in Education is working with many Independent Software Vendors (ISV) and System Integrators (SI) to bring solutions for common education infrastructure challenges like storage, disaster recovery, archiving and content delivery.

## IV. CONCLUSION

By means of providing infrastructure, platform and software as service there are lot of benefits for the educational institution. Educational institution will be benefited by freeing IT Staff from a maintenance, updates and minimal amount of software support. The students' expectations can be satisfied with the rising demand for the latest technology on the campus. Sharing content is as simple as granting someone access, which facilitates collaboration without having to transfer files or worry about software compatibility. The functionality available through Cloud Services is sufficient for the needs of most users, who have access to their files and related software any place they have a computer and an Internet connection. In addition, Cloud Services can work with existing single sign-on programs, and a hardware failure becomes less of a concern. Organizations like Microsoft, Google and Amazon are providing grants and free access for Universities, Colleges, Researcher and students and the educational institutions can use the services with less effort.